# Concept Catalyst: Exploring Scrutable Interfaces to Structure K-12 Teacher Interactions with Generative AI

Gennie Mansi [1], Sunni Newton [2], Roxanne Moore [2], Meltem Alemdar [2], Mark Riedl [1]
[1] Georgia Institute of Technology, School of Interactive Computing
[2] Georgia Institute of Technology, Center for Education Integrating Science, Mathematics, and Computing (CEISMC)

**Purpose:**
This paper explores how to align AI-based tools with teachers' classroom needs by using *scrutable* interfaces—interfaces that link an easily manipulable knowledge representation to an underlying AI model, so users can change the system's outputs without understanding its details. It provides an in-depth discussion and example of a scrutable interface that structures teachers' interactions with generative AI. This study aims to expand how and where scrutable interfaces are used in AI-based tools to support teachers who have not been historically targeted in the design of scrutable systems.

**Design/Methodology/Approach:**
This paper presents the design and evaluation of Concept Catalyst, an AI-based tool with a scrutable interface, created to support teachers' reflection while using generative AI for curriculum development. It presents the findings from an exploratory study using Wizard-of-Oz testing with middle and high school engineering teachers, resulting in 10 depth interviews lasting 55 minutes on average. Screen/audio recordings and the classroom content teachers produced during the session were also collected.

**Findings:**
The paper provides empirical insights about how scrutable interfaces can positively structure teachers' interactions with generative AI models when creating classroom content. Findings suggest that scrutable interfaces can help teachers reflect on their teaching practices while improving efficacy, efficiency, and motivation when using AI.

**What is original/value of the paper:**
This paper explores an identified need to support teachers' classroom practices and needs when using generative AI. It extends the consideration of scrutable interfaces in two ways: to support teachers as users (not just students) and to structure interactions with generative AI models.

## Introduction

Teachers write learning scaffolds to help students build on previous knowledge and practice new skills. To create learning scaffolds, teachers draw on various teaching and professional experiences to create questions, cues, prompts, and other supports to help students systematically master concepts (Nadelson et al., 2014). While teachers recognize adapting and creating learning scaffolds as critical and valuable, they've also described the intensive time and effort needed to write and tailor scaffolds for different student needs (Coppola et al., 2015; Johnson et al., 2017).

Generative artificial intelligence (AI)—that is, AI that produces content—could help teachers meet different pedagogical demands. While students' use of generative AI is debated (Corp & Revelle, 2023), teachers' use of generative AI, such as ChatGPT, has the potential to decrease their workloads (Kasneci et al., 2023; Sok & Heng, 2023) by providing a starting point for brainstorming and writing lesson plans (Sok & Heng, 2023). For example, it could support the creation of classroom content and scaffolds to help students practice concepts or skills. However, many teachers are not experienced in how to best use generative AI to effectively edit or create class content (Corp & Revelle, 2023).

As teachers use generative AI, it is important to support their reflective practices in which they systematically integrate their teaching experiences and knowledge to shape classroom content, including scaffolds. Reflective practices help teachers draw on their teaching and professional experiences (Nadelson et al., 2014) to create and contextualize new curriculum and projects (Chiu et al., 2021). Through their reflective practices, teachers also enhance their perspective of learners (Radhakrishnan et al., 2021), helping them to tailor content to student needs and improve on weaknesses of their teaching (Mesutoglu & Baran, 2020).

Scrutable interfaces may help teachers leverage generative AI technology while supporting their reflective practices. Scrutable interfaces or learning systems use a knowledge representation, such as a concept map, to structure users' interactions with an underlying AI model. Critically, scrutable systems don't require users to understand how the underlying AI model works to change the AI system output. Instead, they support learning by using the knowledge representation as a form of feedback (Khosravi et al., 2022): as users learn and change their actions, they alter the representation, impacting the outputs from the AI (Conati et al., 2018; Kay, 2001) and supporting metacognitive processes of self-monitoring, reflection, and planning (Abdi et al., 2020; Bull & Kay, 2013).

Although scrutable interfaces seem promising, there are several gaps in understanding how to use them to structure teachers' interactions with generative AI when creating classroom content. First, prior work has primarily developed scrutable systems for students as learners, not teachers (Bodily et al., 2018). Additionally, scrutable interfaces have not been widely considered for "black box" models, such as those used in generative AI tools. Instead, scrutable systems have traditionally focused on "opening up" an underlying student model by displaying a visualization representing the computer's key information about their knowledge that impacts outputs. Students interact with the visualization to understand and change what is displayed, subsequently impacting the model's outputs (Bodily et al., 2018). Consequently, more work is needed to explore the design and impact of scrutable interfaces on teachers' uses of generative AI technology.

In this paper, we explore the use of a scrutable interface to support teachers' reflective practices in the context of a generative AI-based tool for K-12 engineering teachers to create classroom content. K-12 engineering teachers' reflective practices are a core part of how they tailor learning scaffolds to help students better document their work, review prior design decisions, and integrate class discussions into their engineering solutions. We present our design and evaluation of **Concept Catalyst** as a practical example of how a scrutable interface can bridge engineering teachers' classroom needs and generative AI abilities. Concept Catalyst is a generative AI-based co-writing tool that allows engineering teachers to create learning scaffolds to engage students' critical thinking skills (see Figure 2). Concept Catalyst's interface integrates into teachers' classroom practices, supporting their reflection while enabling them to more easily create scaffolds.

Through our work, we contribute a practical example and insights about how to structure teachers' interactions with generative AI models to support their classroom and metacognitive needs. We present our findings from a formative study with middle and high school engineering teachers ($n$ = 10). We draw on their experiences to explore and extend the use of scrutable systems to underlying, generative AI models. We argue that scrutable interfaces should be more widely considered for generative AI to serve as channels for content generation and feedback while supporting users' reflective practices.

## Background and related works

First, we situate our work by reviewing engineering teachers' classroom needs and their needs from generative AI technologies. We also review prior work on scrutable interfaces, especially their use of knowledge representations. Then, we explain how we grounded our design decisions for Concept Catalyst in the context of engineering classrooms to create a scrutable learning system that encourages reflection.

## Context: K-12 engineering teachers' classroom needs

K-12 engineering teachers focus on teaching students the Engineering Design Process (EDP), a systematic, iterative process used by engineers to solve problems (Moore et al., 2016). While there are many published versions of the EDP, there is no universally accepted sequence or semantics that describes the process (Moore et al., 2016). Consequently, teachers vary in how they present the EDP's steps, depth of process, structure (linear or circular), and terminology (Moore et al., 2016; O'Donnell et al., 2002; Singhose & Donnell, 2009). However, all models of the EDP are characterized by overarching concepts centered on systematic, iterative inquiry that position the process as a central part of engineering. Through their use of the EDP, teachers help students learn critical thinking skills as they apply different math and science concepts to engineering projects.

K-12 engineering teachers coordinate multiple paths for student learning. First, they engage students through hands-on projects, providing critiques to encourage student creativity, motivation, and problem solving (Alemdar et al., 2018; Moskal et al., 2007). Teachers give students a *design challenge*, a written description of an engineering task or problem that they must address. Students learn by developing and evolving physical artifacts as solutions to design challenges, while teachers offer critiques and troubleshoot challenges (Alemdar et al., 2018; Moskal et al., 2007).

Second, teachers develop students' critical thinking and iterative design skills by requiring written documentation of their reasoning for their solutions, which allows students to review and learn from prior design decisions (Gale et al., 2019; Moore et al., 2016). Engineering teachers leverage writing as a key epistemic practice to develop the "habits of the mind" (Gale et al., 2019) at the heart of learning the EDP (Mangold & Robinson, 2013) and engineering as a discipline (Kelly & Cunningham, 2019). Students develop iterative approaches to addressing design problems by writing documentation—tables, graphs, diagrams, and other written references—that helps them reflect on discussions, trade-offs between designs, and prototype failures (Gale et al., 2019; Moore et al., 2016).

To help contextualize content and guide students' writing (Moore et al., 2016), engineering teachers write extra questions that they incorporate into class documents and discussions (Douglas & Chiu, 2012; Mangold & Robinson, 2013). Students are often unaccustomed to writing in engineering classrooms (Moore et al., 2016) and may struggle to connect their writing to improvements in their final prototype (Moore et al., 2016). So teachers write questions as learning scaffolds to guide their inquiry (Douglas & Chiu, 2012), provide additional practice (Douglas & Chiu, 2012), and encourage creativity (Mangold & Robinson, 2013). For example, to help students brainstorm design requirements, teachers may provide questions guiding students to consider different stakeholders' needs and to evaluate their proposed solutions considering those needs.

## Importance of reflective practices for scaffolding student learning

Engineering teachers exercise a high degree of creativity and problem solving through their multi-faceted engagement with students (Belghith et al. "Problem-Solving", 2023). Teachers draw on various teaching and professional experiences (Nadelson et al., 2014) to create and contextualize new design challenges (Chiu et al., 2021). Creating new design challenges not only involves researching and identifying a new problem that will be tractable and engaging for students but also entails creating new written scaffolds to help students document their work. Even after developing new design challenges, teachers work to tailor scaffolds to students' needs to guide their inquiry and provide additional practice (Douglas & Chiu, 2012).

Teachers' reflective practices, motivations, understanding of the EDP, and practical constraints around assessments are central to how they teach the EDP (Belghith et al. "Problem-Solving", 2023). Reflective practices—in which teachers are guided to systematically integrate their teaching experiences and knowledge—can help improve their understanding of engineering content (Radhakrishnan et al., 2021), discover strengths and weaknesses of their teaching (Mesutoglu & Baran, 2020), revise lesson plans (Mesutoglu & Baran, 2020), and enhance their perspective of learners (Radhakrishnan et al., 2021). Teachers' reflective practices also help them draw on various teaching and professional experiences (Nadelson et al., 2014) to create and contextualize new curriculum and projects (Chiu et al., 2021).

While teachers recognize adapting and creating scaffolds as critical and valuable, they've also described the intensive time and effort needed to write and tailor scaffolds for different student needs (Coppola et al., 2015; Johnson et al., 2017). Scaffolds need to be written from scratch or tailored to different student needs, redirecting teachers' energy away from hands-on engagement (Coppola et al., 2015; Johnson et al., 2017). They do this in addition to altering their curriculum to match frequently changing nation, state, and school-district learning standards (Alemdar et al., 2017; Belghith et al. "Examining Hard", 2023; Mesutoglu & Baran, 2020), leading to significant cognitive and pedagogical loads (Belghith et al. "Problem-Solving", 2023; Belghith et al. "Examining Hard", 2023).

## Engineering classroom needs & generative AI

Generative AI technologies hold the potential to support teacher activities, including lesson planning and assessment creation, thus decreasing teacher workload (Kasneci et al., 2023; Sok & Heng, 2023). Educators debate students' uses of generative AI (Corp & Revelle, 2023), such as ChatGPT, an issue that is amplified in engineering education where students may use these technologies to avoid writing documentation, hindering the development of their critical thinking skills. However, teachers have expressed interest in using generative AI tools in their own work for lesson planning and assessment creation (Kasneci et al., 2023; Sok & Heng, 2023). Prior work has shown that teachers can benefit from the use of generative AI, developing their own critical thinking skills as they assess and adapt suggested lesson plans for their classrooms (van den Berg & du Plessis, 2023).

Generative AI technologies, such as Large Language Models, are often not reliable enough in a fully automated capacity (Schneider et al., 2024), but they can be effectively used with teacher oversight (Nagy et al., 2023). But teachers often don't understand how to effectively prompt generative AI, and there's a recognized need to train teachers on its use (Corp & Revelle, 2023). While prior work has demonstrated that interfaces of generative AI systems can help teachers more easily express themselves (Calo & MacLellan, 2024), it is unclear how to structure interactions with generative AI to encourage teachers' reflective and creative practices in engineering classrooms.

### Scrutable learning systems & knowledge representations to structure AI interactions

Scrutable learning systems have leveraged concept maps (Mabbott & Bull, 2006) and other graphic knowledge representations (Conejo et al., 2012) to help learners understand, reflect on, and change outputs when interacting with an underlying AI system (Conati et al., 2018). Concept maps, knowledge graphs, and tree diagrams are all structured knowledge representations that can support the integration of new information with prior knowledge (Moreno & Mayer, 2007), a key aspect of reflection shown to help improve teacher practices (Ainsworth, 2006; Ardito, 2022; Robinson & Kiewra, 1995). Concept mapping has been used to improve comprehension and reflection (Liu et al., 2018) and integrate new information with existing mental models (Subramonyam et al., 2020). Similarly, Promptiverse (Lee et al., 2022) asked educators to create knowledge graphs which an AI system used to write learning scaffolds—a process that improved the quality of scaffolds while helping educators better understand the content they were teaching.

## Designing the Concept Catalyst

In this study, we present our design and evaluation of an AI-based tool to explore how knowledge representations can productively structure teachers' uses of generative AI. We created Concept Catalyst as a *scrutable* interface. It is an AI-based tool that integrates a concept map into its interface to align several technologies with teachers' needs and classroom practices. As summarized in Figure 1, the concept map plays a critical role in meeting needs in three areas highlighted by prior work: (1) eliciting teacher reflection on their own practices; (2) generating scaffolding prompts in a time-efficient manner; and (3) allowing teacher-driven flexibility of content.

**Figure 1:** *How the Concept Catalyst Meets Engineering Classroom Needs*
A summary of the technical methods employed to meet the various engineering classroom needs identified. We identified three kinds of classroom needs: needs around engineering pedagogy, needs around AI use, and needs around reflective practices. The corresponding background section is noted.

| Teacher Needs | Technical Methods of Support | | |
|---|---|---|---|
| | Generative AI (ChatGPT)[1] | Knowledge Representations [2] | Scrutable Interfaces [2] |
| ***Engineering teachers' needs in pedagogy*** | | | |
| Teachers vary in steps, terminology, and projects used to teach the EDP | | Knowledge graphs have been shown to help in flexibly structuring knowledge for scaffold generation | |
| Scaffolds support students' documentation practices, but are challenging to write | Generative AI has been explored for lesson planning and other teacher activities | | |
| ***Engineering teachers' needs around reflective practices*** | | | |
| Reflection can elicit teachers' wealth of prior experiences | | | Scrutable interfaces can structure reflection and help represent user knowledge |
| Reflective practices can improve teachers' understanding of their content and students' perspectives | | Knowledge representations such as concept maps can help elicit reflection, supporting improved teaching methods and better class content | Scrutable systems help users develop metacognitive skills around reflection, planning, and self-monitoring |
| Reflective practices can help teachers improve how they teach | | | |
| ***Engineering teachers' needs around generative AI use*** | | | |
| Teachers don't always know how to use generative AI to get the response they want | | | Scrutable interfaces can help people understand how to interact with an AI system without knowing how the system works |
| Teachers want to use AI systems in a way that supports students while easing their own cognitive and pedagogical load | Generative AI can help write content for key classroom activities, including students' documentation practices | | |

[1]See Background section on "Engineering classroom needs & generative AI"; [2] See Background section on "Scrutable learning systems & knowledge representations to structure AI interactions"

To facilitate reflection and flexibility of content, teachers start with a paragraph describing the design challenge. Concept Catalyst prompts the teacher to highlight important concepts in the paragraph (top left, Figure 2). The paragraph addresses the problem, general design needs, and clients—mirroring *what they would expect students to do* if they were writing a paragraph about the problem statement. Because teachers vary in how they present the EDP (see Section 2), we allow teachers to use a student's response or their vision of what students should write. Alternatively, the computer can generate this text if the teacher inputs the design challenge provided to students.

Concept Catalyst then prompts teachers to identify how each highlighted concept relates to other concepts they already highlighted and creates a graphic visual showing the hierarchical relationships between the concepts students need to address (bottom left, Figure 2). The teacher can create as complex of a concept map as they want, and they can remove, edit, and add any concepts, even those not shown in the sample paragraph.

Once the teacher has finished the concept map, they select as many concepts as they want from the visualization for the AI system to write scaffolding questions. The tool inserts the selected concepts into a pre-written prompt for ChatGPT to generate scaffolds, which are returned to the teacher via the tool's interface (right, Figure 2). After the prompts are generated, we ensure the teacher maintains final control over class content by allowing them to save, edit, regenerate, or remove scaffolds created. Teachers can create as many sets of questions as they would like, and they can disseminate them as they wish—in their own notes, in slides as discussion questions, worksheets, or as requirements for students' documentation. Thus, content is flexibly generated according to teacher's priorities and practices.

While Concept Catalyst does not allow teachers to scrutinize the inner workings of the GPT model, it is still *scrutable* in the sense that it helps them understand, reflect on, and directly edit the inputs to the GPT model to achieve their desired outcomes. Further, the concept map in Concept Catalyst mirrors the role of student-facing visualizations in traditional scrutable systems. Because it is developed through reflection and their normal classroom practices while reviewing student responses, the concept map also effectively reflects teachers' current state of knowledge about the EDP and design challenge. As teachers reflect, they iterate on their concept map, impacting the outputs of the GPT model, just as the display in a traditional scrutable system updates based on interactions in learning activities.

Critically, this process also places teachers in control of the scaffolds' content at three critical points—(1) the kinds of concepts included in the concept map; (2) which concepts are selected for each set of scaffolds; and (3) how scaffolds are disseminated—which allows the teacher to leverage generative AI while retaining the power of how they guide their students to learn the EDP. Figure 2 shows screenshots of Concept Catalyst with an example concept map and scaffolds generated by the computer.

## Methods

Because our goal is to understand how scrutable interfaces and knowledge representations can structure teachers' reflective practices while using generative AI, we present our findings from Wizard-of-Oz (WoZ) usability testing with teachers. In a WoZ study, users or participants interact with systems that seem autonomous but are covertly operated by a human (Yang et al., 2026). WoZ studies can be leveraged in exploratory phases of design for interaction depth and benchmarking, helping researchers establish what a tool or system should do and why (Yang et al., 2026).

In keeping with best practices and epistemic goals, we conducted remote, WoZ studies with middle and high-school teachers to understand their uses of the knowledge representations in Concept Catalyst when generating scaffolds for actual design challenges. In our interviews, we explained the purpose and motivation of Concept Catalyst. Then, we presented an overview of the workflow for using the tool. Since we wanted to elicit their feedback on the scaffolds and interface features before building out the system, we provided teachers a mock-up of Concept Catalyst created using Miro (https://miro.com/)—an online collaborative visual workspace—and walked through an example of how to use the interface.

Once their questions were answered, teachers were asked to use Concept Catalyst to create scaffolds for the same design challenge as the researcher "acted out" the part of the computer, effectively mirroring in person paper prototyping. The teacher directed the researcher to create entries for the concept map and select concepts to include in the prompts. The researcher used ChatGPT in the background out of view of the participant to write the scaffolding prompts using a template that remained the same for all interviews (see Supplemental Materials for template). The generated scaffolds were then copied and pasted back into the teacher-facing mock-up, where the teacher could edit and provide verbal feedback to the researcher about the prompts.

We recruited 10 teachers for WoZ testing. We recruited participating teachers through online messaging and contacts of the research team, followed by snowball sampling. Interested participants filled out a brief survey to ensure they met the inclusion criteria: of full-time employment as a high school (Grades 9-12; Ages 14-18) or middle school (Grades 6-8; Ages 11-13) engineering teacher and at least 1 year of teaching experience.

**Figure 2:** *Concept Catalyst Example*
Teachers use Concept Catalyst's interface to create a concept map reflecting the most important parts of an engineering design challenge. To support ideation of concepts, we have teachers highlight phrases in a summary paragraph of the design challenge. They can enter a summary paragraph, or the computer can generate it. Teachers highlight and organize the most important concepts in a concept map. Then they select concepts for the AI to use to write learning scaffolds for their students. The computer outputs its response to the interface where teachers can edit the prompts, ask the AI to regenerate the prompts, or clear the prompts entirely.

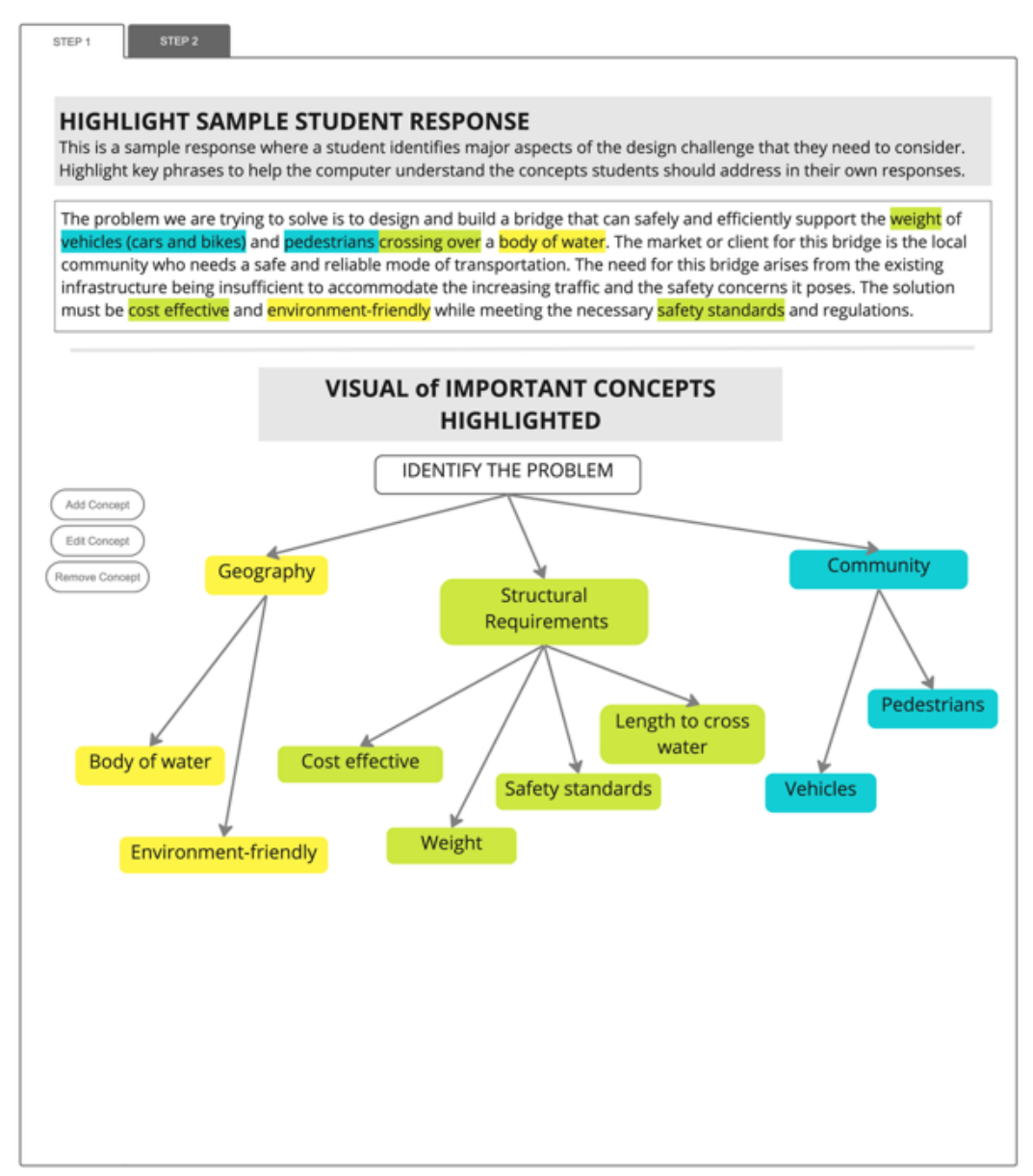

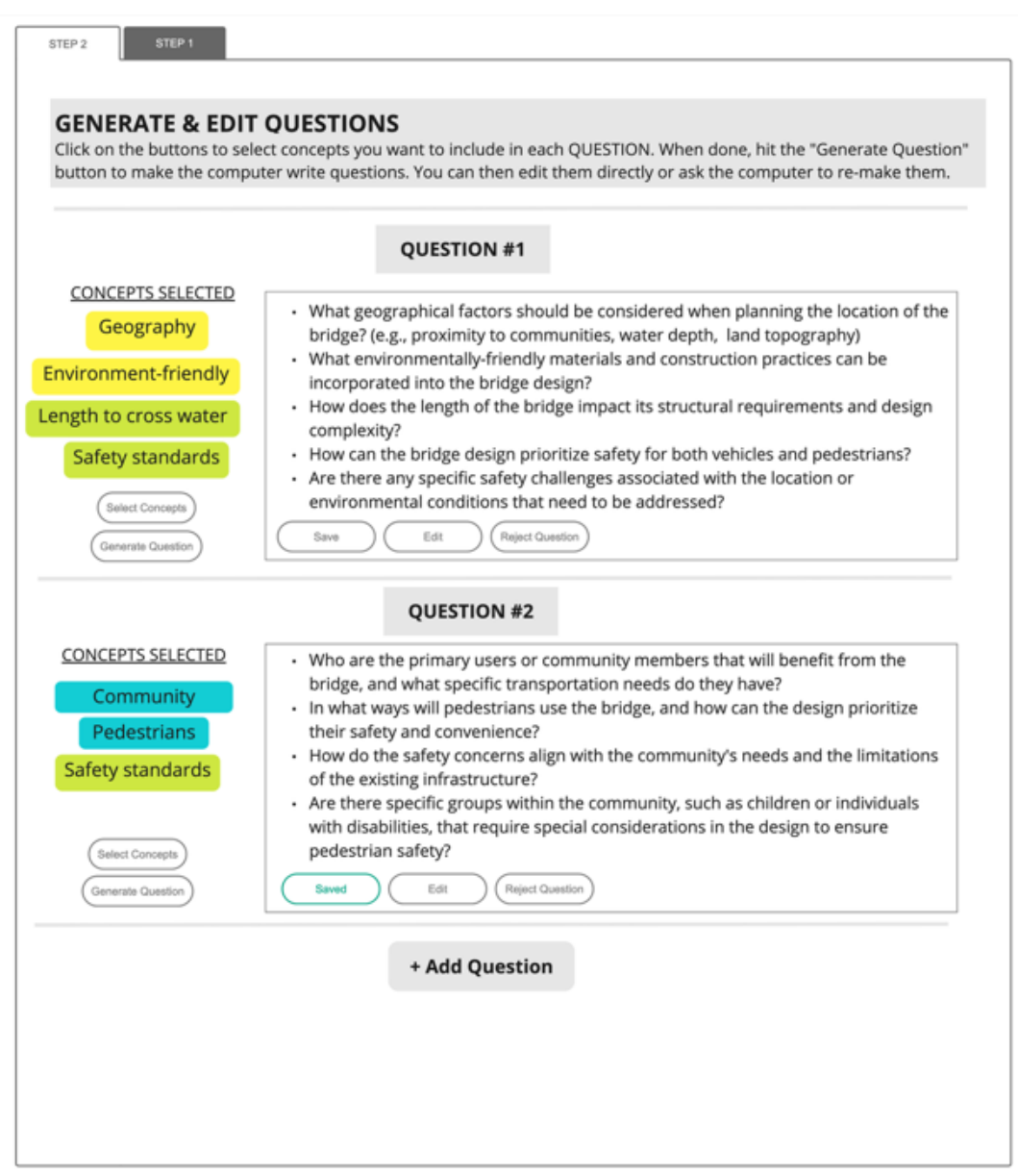

We recruited teachers with a diversity and breadth of teaching experiences. Two teachers had 1-3 years of teaching experience; two had 4-9 years of experience; one had 10-15 years of experience; and five had over 15 years of teaching experience. Five of the teachers had industry experience in engineering-related fields. One teacher had significant informal engineering experience. Interviews lasted approximately 55 minutes on average. An institutional ethics review board approved this study, and we obtained informed consent before all interviews.

## Analysis and results

To understand how scrutable interfaces and knowledge representations structured teachers' reflective practices while using generative AI, we transcribed and analyzed the interviews using affinity diagramming, a technique in which quotes are sorted by theme (LUMA Institute, 2012). Affinity diagramming is often used to understand the usability of a tool (Kaufmann, 2005), a focus of our study. The first author conducted an initial analysis of themes, which were discussed by the rest of the research team until consensus was reached.

We uncovered key findings around (a) teachers' perceptions of the efficiency, diversity, and quality of the scaffolds they created; and (b) the teachers' experience using the tool, including how we supported their reflection. In this section, we present these key findings as a case study of how concept maps can support teachers' reflection while using generative AI. In the following section, we describe how these findings open new directions for work designing scrutable systems to support productive uses of generative AI.

### Diversity, quality & efficiency of scaffolds

#### Efficiency

All teachers agreed Concept Catalyst would reduce their cognitive load because "instead of having to create your own questions the AI could generate those questions for you" (P3). Several teachers (P1, P7) indicated their preference for Concept Catalyst over other AI-based tools because it helped them more accurately prompt ChatGPT. This included making adaptations for frequently changing school district requirements. P7, P8, P9, and P10 also voiced that the interface's structure helped them quickly generate content that they could more easily adjust, lowering barriers to teacher innovation and freeing them to work with students and "do what you came into the profession to do" (P7).

#### Quality

All teachers voiced that the questions were interesting and relevant to the design challenge. They felt questions were similar in quality to those they write. They voiced that the questions could support students' critical thinking by forcing them to write more and giving them more ideas to write about in their responses. Some teachers (P2-P4, P6) felt questions were written in "fancy talk" (P2) or were too long (P3) and needed modification. Still, teachers preferred our tool to output more content rather than less, requesting interface features such as color-coding (P3, P7, P10) and organization methods (P2) to help them "massage out" what was needed.

#### Diversity

Teachers observed Concept Catalyst matched their content focuses, such as specific definitions of "criteria" and "constraints" (P1, P4, P9), incorporating soft skills such as teamwork (P1), tailoring content for student interests (P6), and prompting students to make connections to personal experiences (P8). Teachers created scaffolds to engage students in higher order thinking (P7) and through re-framing the same topic in multiple ways (P6). They also created scaffolds for various in class activities, including verbal discussions (P4, P8, P9), homework assignments (P4, P8, P9), assessments (P8, P10), and longer projects (P2). Finally, teachers created scaffolds to match their teaching styles. While P2 and P4 sharply edited the scaffolds, so they could present them in a more scripted way, other teachers preferred to "create a library of prompts [scaffolds]" (P1), so they could "feed off" (P8) students' energy.

### Supporting teacher reflection

We found that Concept Catalyst supported teachers' reflection around the engineering design process and their teaching practices throughout the writing process. Initially, most teachers were skeptical of using AI at all. Several teachers were also unsure of what to add to the concept map (P1, P2, P5-P9), but reading the sample paragraph helped them remember their teaching experiences. For example, P9 read the sample response and added "construction cost" as a concept under "materials" because "in class they (the students) would have to choose how much material they're going to use and how much glue and tape and things for the construction". In this way, the sample response gave teachers a place to start generating and refining their ideas when they had an idea but "not necessarily the words for it" (P5).

Prior to writing any content, Concept Catalyst's design helped teachers synthesize their thoughts and organize what students should learn from an engineering design challenge in two ways. The *concept map* helped teachers **iterate on what students should address in their writing**. P2, P6, P7, P9 and P10 all continually revised the concept map, describing it as a "mind map" (P9) that helped them "outline what (they) want" (P2) students to address by helping them visualize and iterate on how the concepts connected. The *concept map* also helped teachers **think of ways to lesson plan within and across classes**. P4, P8, P9, and P10 mentioned using the concept map to figure out a group of topics they could explore with students in a lesson depending on their interests.

After the concept map was finalized, and teachers had selected concepts for the AI-system to write about, Concept Catalyst continued to support teacher reflection and iteration on pedagogy. P4, P6, P7, and P9 all highlighted that the *learning scaffolds* written by the AI system **helped them think of new content to include in their classroom**. For example, when a question was generated about building an aesthetic bridge, P4 expressed appreciation that it presented "a new way to look at things". Concept Catalyst's outputs inspired both P6 and P7 with ideas for new activities, and P7 emphasized the unique benefit of using Concept Catalyst to think about things that "Google would never give you." Further, reading through the output from Concept Catalyst helped teachers reflect and iterate on their classroom practices, including the kinds of questions they asked (P1, P2, P4, P8); the design of existing activities in unrelated design challenges (P1, P2); connections to student interests (P6); and updates to in-class assessments (P6, P8). Several (P5, P6, P8, P9) credited the benefits of Concept Catalyst to the way it helped them practice the EDP, explaining they wanted their students to use it too.

## Discussion

We used prior work on K-12 engineering classrooms (see Section 2) to identify the common needs of teachers and students and examine where generative AI could best support their activities. While generative AI could reduce teachers' cognitive and pedagogical load by writing scaffolds to support student documentation, teachers have trouble reflecting on their experiences and appropriately prompting generative AI (Corp & Revelle, 2023). To address this we created Concept Catalyst, a ChatGPT-based system through which teachers create a concept

map—a visual representation of the relationships between concepts—which is then used to write learning scaffolds for students.

Concept Catalyst was designed as a *scrutable* interface, providing an easily manipulable knowledge representation linked to an underlying AI system, so users could change the outputs of the system without understanding the details. By doing this, we aimed to help teachers more easily leverage ChatGPT while reflecting on their teaching practices (Ardito, 2022; Khosravi et al., 2022)—recalling specific classroom experiences (Moreno & Mayer, 2007), better understanding the concepts they're teaching (Ardito, 2022; Lee et al., 2022), and making targeted changes to their teaching (Cornford, 2002; Moreno & Mayer, 2007).

Our findings from 10 WoZ studies with engineering teachers demonstrate how scrutable interfaces that leverage knowledge representations, such as concept maps, can positively structure teacher reflection while they use generative AI. Teachers deeply appreciated the efficiency, quality, and diversity of scaffolds they could write with Concept Catalyst. More importantly, Concept Catalyst deepened their personal engagement with the EDP. P1 stated that the system was an effective tool because it helped them cognitively practice the skills and processes that they wanted students to practice, a main reason they could innovate new ways of structuring the classroom. This was mirrored in the other teachers' activities as they actively reviewed the concept map and scaffolds, leading teachers to reflect on their teaching practices, class experiences, students, and industry experiences as they adapted their teaching and innovated new activities.

The benefits we observed hold design implications for what it means to "explain" an AI model and support positive user outcomes. Previously, explainable AI models in education placed a high premium in allowing people to examine the details of a model. While Concept Catalyst does not allow teachers to scrutinize the inner workings of the GPT model, it is still *scrutable* in the sense that it helps them understand, reflect on, and directly edit the inputs to the GPT model to achieve their desired outcomes. The teachers in our study not only felt encouraged to use AI via Concept Catalyst, many wanted their students to use it, too—a significant shift from the initial hesitancy we observed at the start of the interview and that is also reported throughout literature (Kasneci et al., 2023; Lo 2023; Mosaiyebzadeh et al., 2023). Consequently, we broaden the usage of the term *scrutable* to include similar systems to ours.

While scrutable learning systems have historically targeted students (Khosravi et al., 2022), Concept Catalyst demonstrates the benefits for others in the classroom too. Often, developers turn to AI explanations—descriptions of an underlying model—to help users interact with and change AI systems. However, designers need to consider when a *scrutable* interface may better bridge a sociotechnical gap than an *explainable* one. The importance of learning the abstract thought process of the EDP in the classroom context was facilitated through a knowledge representation in a *scrutable* interface, engaging higher order metacognitive skills alongside use of the AI. Considering the social needs of users and the technical capabilities of generative AI to determine if a scrutable or explainable system better fits the context, can significantly improve users' outcomes, supporting more than just better AI outputs.

## Limitations

Engineering education varies widely across classrooms. A broader sample of educators, including specific efforts for recruiting diverse perspectives in terms of geography, culture, and socioeconomic status, would enhance our findings. Additionally, our case study with teachers tested Concept Catalyst, and we had all teachers build concept maps for the same, well-known, bridge design challenge to more easily support comparisons across teacher interactions. More work is needed to understand how their interactions may change when writing for their own projects. Finally, we didn't interview any students for their perceptions of the questions. Several teachers (P5, P6, P8, P9) also wanted a student-facing version of the Concept Catalyst to help teach problem solving. Future work should explore how the current tool could be adapted for student use, including collaborative work and different reading abilities.

## Conclusions

K-12 engineering teachers new content, writing scaffolds for students, and hands-on engagement. Our paper contributes an example for designing scrutable interfaces to support teachers' use of generative AI to support classroom activities. We discuss how we developed Concept Catalyst as a scrutable interface to bridge between the pedagogical needs of teachers and the technical abilities of ChatGPT. Concept Catalyst leverages a concept map to help teachers reflect on their experiences and direct the creation of content through ChatGPT without requiring them to understand exactly how the GPT model works. We interviewed middle and high school engineering teachers who gave us positive feedback on the efficiency, quality, and diversity of content they wrote. Concept Catalyst supported teacher reflection that helped them more deeply engage with the EDP while innovating methods for student engagement. We hope future work will develop other kinds of technology that can serve as scrutable interfaces bridging sociotechnical gaps to better fit the needs of educators and students as they grow in their ability to experience, understand, and enjoy both classroom content and AI's abilities.

## Acknowledgements

(Redacted).